\documentclass[]{jd04}    
\usepackage[]{graphicx}

\Title[{\sc Wind3D}]{Three Dimensional Radiative Transfer in Winds of 
Massive Stars: Wind3D} \Author{Lobel$^1$, A.} \Author{Blomme$^1$, R.}
\Address{$^1$Royal Observatory of Belgium,\\
Ringlaan 3, Brussels, B-1180, Belgium \\
}{}

\ShortAuthors{Lobel \& Blomme}


\begin{document}
\maketitle

\begin{abstract}

\centerline{\em Radiative transfer calculations in three dimensions are presented}

\centerline{\em to investigate dynamic 
structures in extended winds of massive stars}.

We discuss the development of the new radiative transfer code {\sc Wind3D}. 
It solves the non-LTE radiative transport problem in moving stellar atmosphere 
models in three geometric dimensions. The code accepts arbitrary 3D velocity fields in Cartesian geometry without assumptions of axial symmetry. {\sc Wind3D} is currently implemented as a fully parallelized (exact) accelerated lambda iteration scheme with a two level atom formulation. The numerical transfer scheme is efficient and very accurate to trace small variations of local velocity gradients on line opacity in strongly scattering dominated extended stellar winds. We investigate the detailed formation of P Cygni line profiles observed in ultraviolet spectra of massive stars. We compute the detailed shape of these resonance lines to model local enhancements of line opacity that can for instance be caused by clumping in supersonically expanding winds. {\sc Wind3D} will be applied to hydrodynamic models to investigate physical properties of discrete absorption line components. 

\end{abstract}

\section{Introduction}

{\sc Wind3D} is a new radiation transport code that computes detailed spectral line profiles formed in the scattering winds of massive stars. The time-independent 
transfer equation is solved following the numerical scheme proposed by Adam (1990) for three-dimensional Cartesian grids. The Fortran code is being developed for high performance computers with parallel processing. It currently runs on the 64-bit 56 CPU compute server of the Royal Observatory of Belgium. Figures 1 to 3 discuss the results of calculations with {\sc Wind3D} performed with $80^{3}$ gridpoints on an equidistant mesh. The code lambda iterates the 3D source function to accuracies better than 1\%. The local mean intensity integral currently sums $80^{2}$ spatial angles for 100 wavelength points of the line profile. The non-LTE 3D transfer equation for the 2-level atom is solved for $640^{3}$ interpolated gridpoints using a Gaussian profile function. {\sc Wind3D} has carefully been load balanced for parallel processing and shows excellent scaling properties.

\section{Co-rotating Interaction Region Modeling}

We test {\sc Wind3D} with parameterized input models of the wind velocity and wind opacity in the supersonically accelerating winds of OB stars. The line source function of the unperturbed wind is computed in the Sobolev approximation. For strong resonance lines formed in the winds of massive hot stars we currently neglect the thermal contribution to the line source function. We consider a beta law for an isothermal wind with 
$\beta$=1 and $R_{\star}$=35 $\rm R_{\odot}$. The {\it smooth} wind assumes a terminal velocity 
of $v_{\infty}$=1600 $\rm km\,s^{-1}$ within the simulation box of 12 $R_{\star}$. The thermal line broadening is set to a small value of 8 $\rm km\,s^{-1}$. The smooth wind is perturbed with parameterized spiraling structures that are wound around the star ({\it Figs. 1-2}). Next the source function in these regions of larger gas density is 3D lambda iterated to equilibrium with the radiation field of the ambient wind. The convergence can be accelerated but our tests show that it is already fast and occurs within 5 to 8 iterations when the azimuthal extension of the CIRs is limited to one $R_{\star}$ at the outer edge. The converged 3D source function is then utilized to solve the transfer problem for equidistant lines of sight in the plane of the equator around the star. Important free parameters for the detailed line profile modeling therefore involve the 3D properties of the spiraling wind structures (the number of arms, curvature, width, height, and density contrast), and the inclination of the observer lines of sight in or above the plane of the equator. These simplified 3D CIR models already provide good comparisons with the observed evolution of DACs in UV spectra of hot stars (e.g. see Fig. 1 of Fullerton et al. 1997). Further tests with dense spherical perturbations of the wind (i.e. a clump in Fig. 3) reveal that a shadow behind the clump can be computed with {\sc Wind3D}, which also appears in the theoretical line profiles ({\it left panel of Fig. 3}).

\begin{figure}[!ht]
  \hspace*{0.5cm}
  \begin{minipage}[h]{5 cm}
    \includegraphics[width=5.5cm]{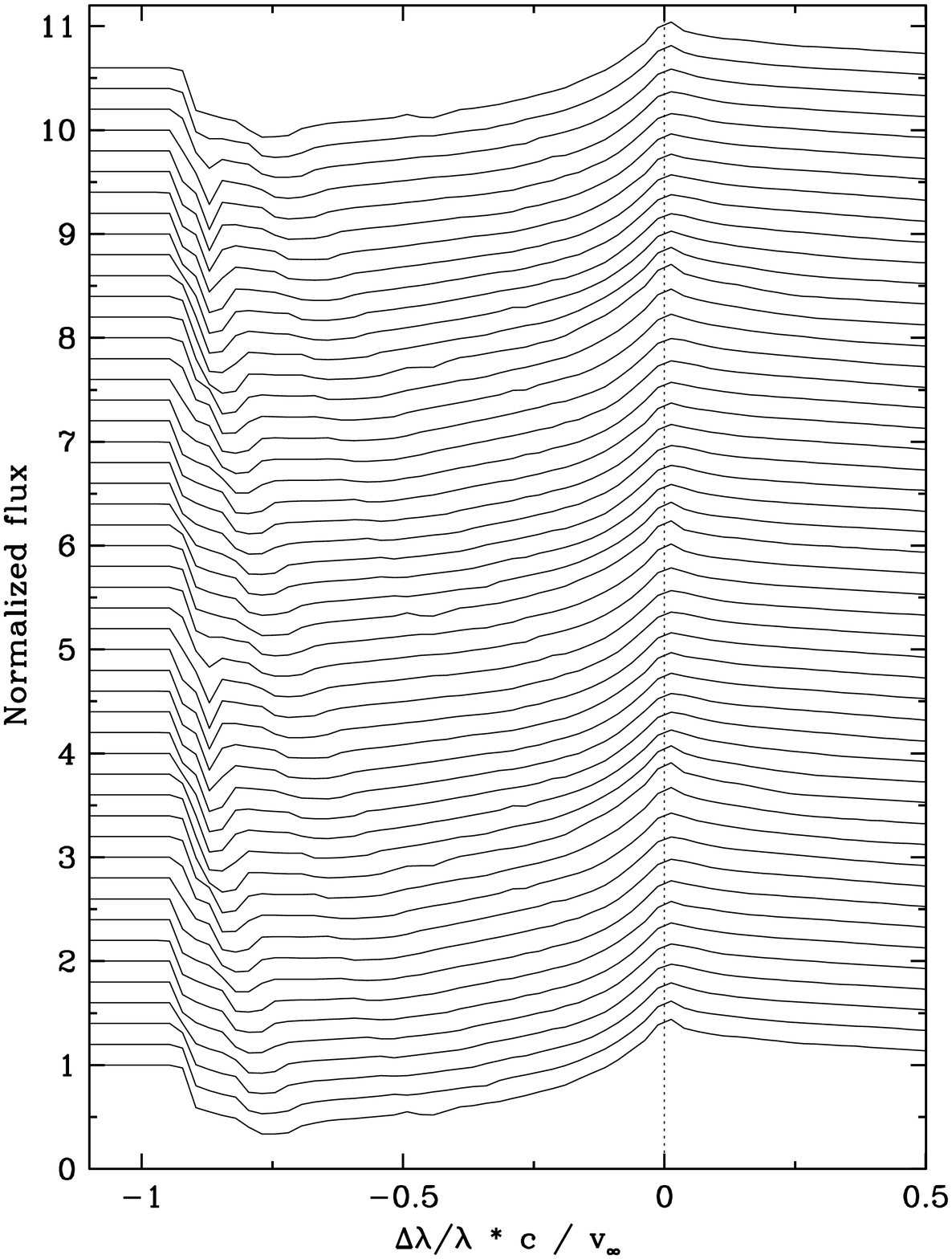}  
  \end{minipage}
  \hspace*{0.5cm} 
  \begin{minipage}[h]{5 cm}
    \includegraphics[width=5.cm]{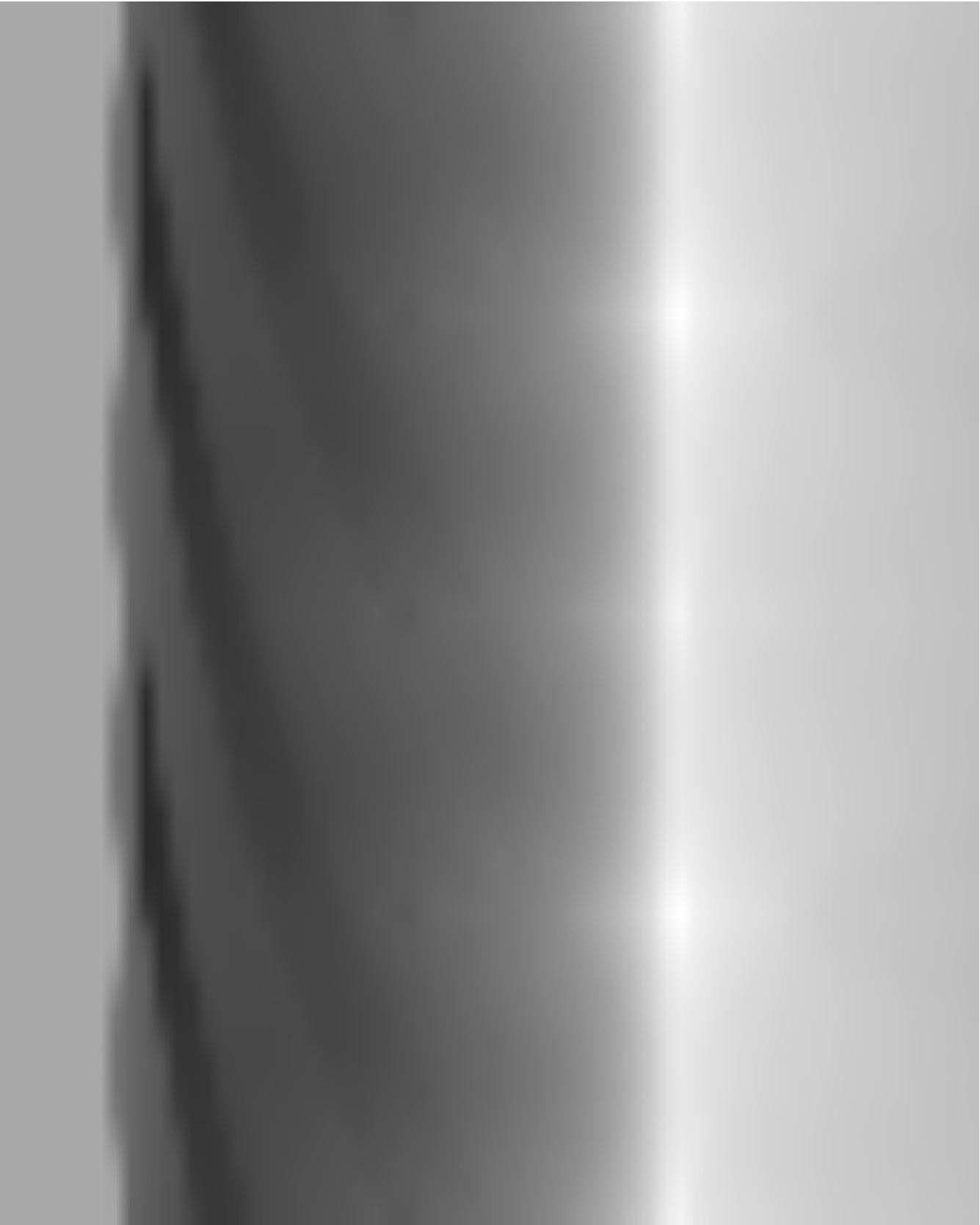}  
  \end{minipage}
  \caption{ Dynamic spectra generated from 3D radiative transfer in the expanding wind of a hot star with two co-rotating interaction regions having a shape of winding density spirals located in the plane of the equator. {\bf Left:} Computed line profiles show two discrete absorption components (DACs) that shift toward shorter wavelengths in the unsaturated absorption trough of the P Cygni line profile (upwards: phases from 0 to 1). {\bf Right:} The same time sequence of line fluxes shown with grayscales. The width of both DACs decreases while drifting toward shorter wavelengths and merging with the violet edge of the P Cygni line profile. The DACs become narrower because the velocity gradient of the expanding wind decreases at larger distances from the surface, while the wind velocity increases to the terminal wind velocity. The line opacity in the co-rotating interaction region (CIR) model is increased by one order of magnitude with respect to the surrounding smooth wind opacity. The radiation line transport with {\sc Wind3D} accounts for the volume of the central hot star and the azimuthal extension of the CIRs perpendicular to the lines of sight. The dynamic spectra are computed for 50 lines of sight in the equator plane around the star.}
  \label{fig12}
\end{figure}

\begin{figure}[!ht]
\centerline{\includegraphics[width=20pc]{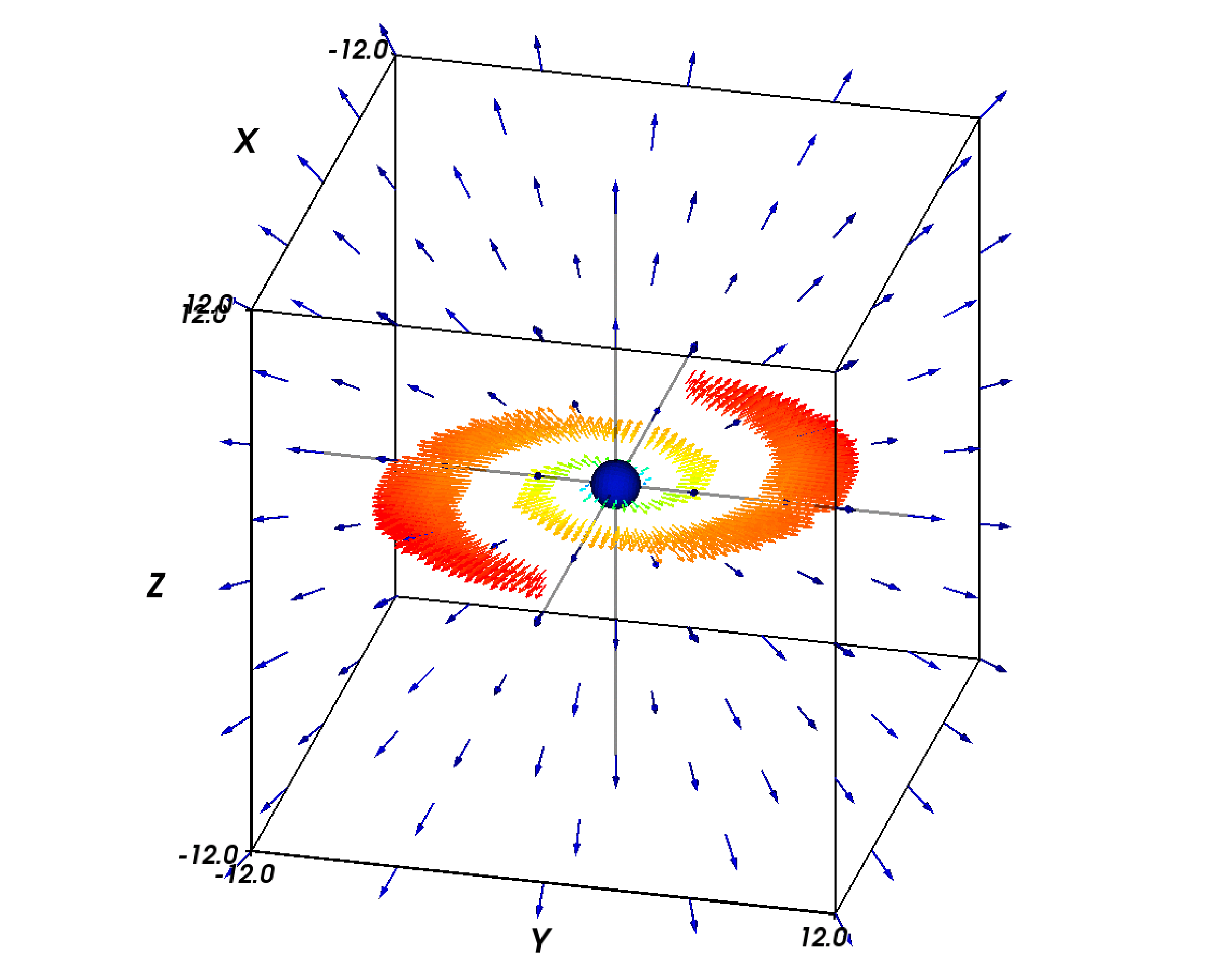}}
\caption{ Schematic drawing of the parameterized wind velocity grid in {\sc Wind3D}. 
The simulation box size is 24 $R_{\star}$ with the hot star at 
box center. The smooth stellar wind is radially symmetric with a beta-power velocity law 
({\it outer blue arrows}). The local velocities inside the CIRs also assume the beta law of the ambient wind, but the velocity vectors are drawn with larger density ({\it inner green to red arrows}). 
The  widths of the CIRs increase 
outwards with an exponential curvature. The outer edges are truncated at a maximum radius of 9 $R_{\star}$. 
The thickness of the CIRs in the equator plane increases outwards to a maximum of one $R_{\star}$. 
The flux contrast of the DACs with respect to scattering wind absorption increases and decreases 
while shifting bluewards ({\it upwards in Fig. 1}) because the DACs narrow (due to the decrease of local 
wind velocity dispersion) and the line opacity (i.e. local gas density) inside the CIRs decreases radially. } 
\label{fig3}
\end{figure}

Fullerton et al. (1997) show in their figure 1 (panel right) the dynamic spectrum of the Si~{\sc iv} $\lambda\lambda$1394 doublet lines observed in the {\it IUE} MEGA Campaign of 1995 of the early B-type Ib supergiant HD~64760. DACs in their figure 1 are recognized as rather narrow absorption features 
observed to appear around $\sim$1 d and $\sim$11 d, drifting towards shorter wavelengths. The broader 
and nearly horizontal absorption features with $\sim$1.2 d periodicity are called {\it modulations} and 
are not considered in our 3D models. The observed time evolution of the DACs is well reproduced with 
the model in Fig. 2. The DACs drift toward larger velocities because the observer probes regions of 
enhanced wind absorption inside one spiral arm at increasing distances from the stellar surface 
while the entire structure rotates through his line of sight. The other (anti-symmetric) spiral arm produces the trailing DAC that drifts almost parallel to the former until it also reaches the terminal wind 
velocity in Fig. 1. The number of spiral arms crossing the line of sight determines the number of DACs 
in a line profile. The rotation period of HD~64760 is 4.8 d, which indicates that the spiral arms 
producing the DACs are wound at least twice around the central star. Other 3D models with thicker 
spiraling density structures extending azimuthally (with much larger opening angles) also produce DACs 
for lines of sight above the equator plane. In the latter case the DACs reveal density enhancements 
inside the spiral arms following the surface of a cone centered around the stellar rotation axis.

\begin{figure}[!ht]
  
  \hspace*{0.1cm}
  \begin{minipage}[b]{5 cm}
    \includegraphics[width=12pc]{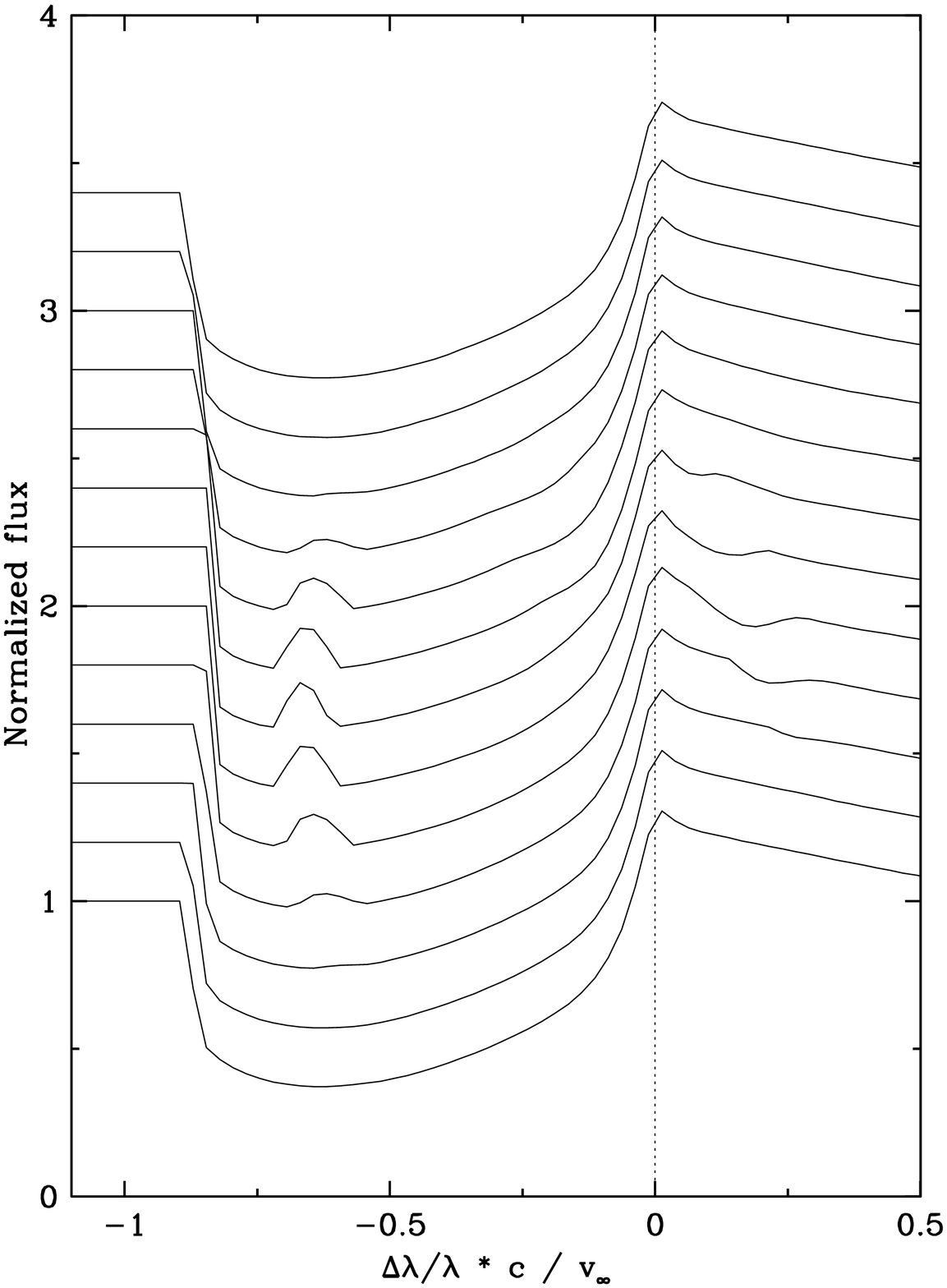}  
  \end{minipage}
   \hspace*{0.1cm} 
  \begin{minipage}[b]{5 cm}
    \includegraphics[width=20pc]{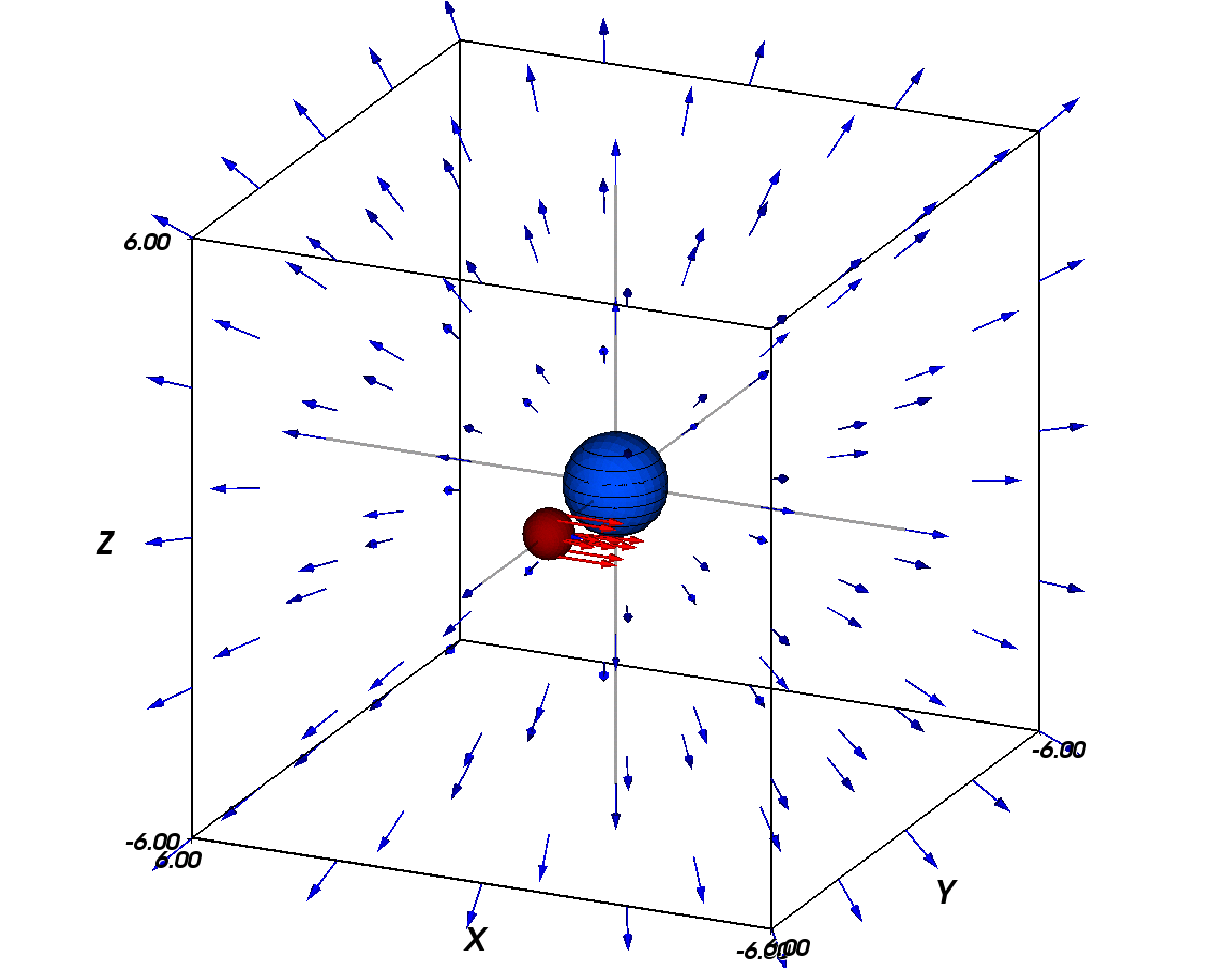}  
  \end{minipage}
  \caption{{\bf Left:} Line profiles computed with {\sc Wind3D} using the 3D model in the right-hand panel. 
  The dynamic spectrum reveals how the absorption portion of the P Cygni profile due to wind scattering 
  in front of the stellar disk decreases when a local opacity enhancement (a spherical clump model) 
  crosses the observer lines of sight. The dynamic spectrum is computed for 13 lines of sight in the plane 
  of the equator over 45 degrees in the front XZ plane of the box. The amount of absorbed line flux decreases
  when the opacity clump passes in front of the stellar disk and partly obscures it. The clump decreases the
  wind scattering volume in the cylinder towards the observer and causes a region of shadow behind the clump
  where photon scattering by the wind is substantially diminished.      
  {\bf Right:} Schematic drawing of the wind velocity model with a spherical clump ({\it small red dot}). 
  The clump has a radius of 1/2 $R_{\star}$ and passes at 3 $R_{\star}$ in front of the central hot star 
  ({\it large blue dot}). 
  The size of the simulation box is 12 $R_{\star}$. The clump moves perpendicular 
  ({\it tangentially drawn arrows}) to the symmetrical radially (beta law) expanding wind ({\it 
  outer blue arrows}). The opacity in the clump is increased 
  with an order of magnitude compared to the ambient wind opacity. The increased opacity of the 
  clump determines the amount of decrease of line absorption computed at $\sim$70\% of the terminal wind 
  velocity shown in the left-hand panel. Other calculations with {\sc Wind3D} show how the line absorption decreases further 
  with larger clump densities or larger clump sizes. The distance of the clump to the star determines the
  velocity position of the flux bump in the P Cygni profile. The position of the flux bump is invariable
  because the stellar wind velocity profile is spherically symmetric and the clump stays at the same 
  distance from the star in the model computed for lines of sight in the plane of the equator. }
  \label{fig45}
\end{figure}

\begin{figure}[!ht]
\centerline{\includegraphics[width=17pc,angle=90]{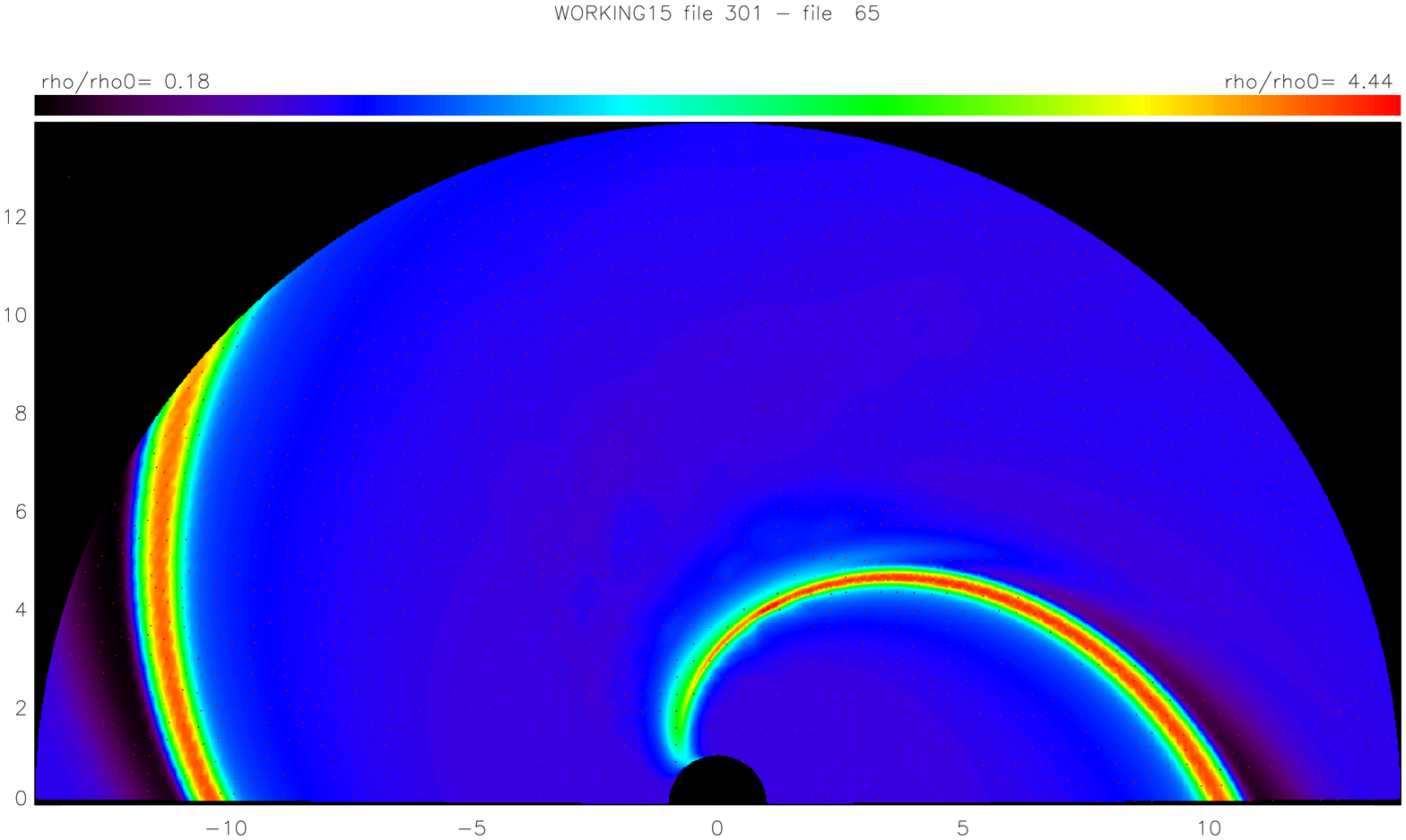}}
\caption{   Hydrodynamic simulation (using the {\sc Zeus} code) of a 2D stellar wind with co-rotating interaction regions. A bright spot on the 
stellar surface gives a greater line force, resulting in an enhanced wind density extending from the stellar surface to 14 $R_{\star}$. The star rotates with 350 $\rm km\,s^{-1}$ at the surface. Three dimensional
hydrodynamic models will be computed with {\sc Zeus-3d} as input for {\sc Wind3D} to investigate the physical properties 
of DACs observed in hot star winds.                 }
\label{fig6}
\end{figure}

\section{Summary and Outlook}

We are developing a three dimensional radiative transfer code {\sc Wind3D} to investigate the physics 
of co-rotating interaction regions and clumping in winds of luminous hot stars. Preliminary tests 
show that simplified models with density spirals around the star well correspond to the properties of discrete absorption profiles observed in UV resonance lines. The code will be further developed for multi-level 
atoms and tested for more complex density and velocity structures in the wind. {\sc Wind3D} will be utilized 
to semi-empirically determine density contrasts from detailed variability observed in the dynamic spectra. 
The parameterized wind models will be replaced with ab-initio hydrodynamic models computed with {\sc Zeus-3d} 
({\it Fig. 4}) to investigate the formation physics of these remarkable wind structures.

\noindent
{\it Acknowledgements:}

This work has been supported by the Belgian 
Federal Science Policy - Terug-keermandaten.

\end{document}